\documentclass[reprint,showpacs,amsmath,amssymb,%
   aip,apl,longbibliography]{revtex4-1}

\usepackage{epsfig}
\usepackage[utf8]{inputenc}
\usepackage{color}

\newcommand{\vg}{\ensuremath{V_\text{g}}}
\newcommand{\vsd}{\ensuremath{V_\text{sd}}}

\newcommand{\didv}{\ensuremath{\text{d}I/\text{d}\vsd}}
\newcommand{\un}[1]{\ensuremath{\,\text{#1}}}

\newcommand{\nel}{\ensuremath{N_{\text{el}}}}

\begin{document}

\title{Liquid-induced damping of mechanical feedback 
effects in single electron tunneling through a suspended carbon nanotube}

\author{D. R. Schmid}
\author{P. L. Stiller}
\author{Ch. Strunk}
\author{A. K. Hüttel}
\email{andreas.huettel@ur.de}

\affiliation{%
Institute for Experimental and Applied Physics, University of
Regensburg, Universitätsstr.\ 31, 93053 Regensburg, Germany
}

\date{\today}

\begin{abstract}
In single electron tunneling through clean, suspended carbon nanotube devices 
at low temperature, distinct switching phenomena have regularly been observed.
These can be explained via strong interaction of single electron tunneling and 
vibrational motion of the nanotube. We present measurements on a highly stable 
nanotube device, subsequently recorded in the vacuum chamber of a dilution 
refrigerator and immersed in the $^{3}$He/$^{4}$He mixture of a second dilution 
refrigerator. The switching phenomena are absent when the sample is kept in 
the viscous liquid, additionally supporting the interpretation of dc-driven 
vibration. Transport measurements in liquid helium can thus be used for 
finite bias spectroscopy where otherwise the mechanical effects would dominate 
the current. \\[0.2cm]
(c) AIP Publishing LLC \hfill {http://dx.doi.org/10.1063/1.4931775}
\end{abstract}

\maketitle

Clean, suspended carbon nanotubes provide an extremely versatile model system. 
As nano-electromechanical beam resonators, they can be tuned over a large 
tension and thereby also frequency range.\cite{nature-sazonova:284, 
nl-witkamp:2904, highq, highqset, pssb-stiller:2518} At cryogenic temperatures, 
very high mechanical quality factors have been observed,\cite{highq, 
nl-island:4564, mosermillion} making the observation of non-trivial 
interaction between single electron charging and the mechanical motion 
possible.\cite{strongcoupling, science-lassagne:1107, prb-meerwaldt:115454, 
Benyamini2014} Both electronic tunneling \cite{prb-meerwaldt:115454, 
prl-ganzhorn:175502} and magnetic induction \cite{magdamping, 
prb-nocera:155435} have been shown to induce damping and thereby reduce the 
effective mechanical quality factor.

In addition, clean carbon nanotubes also provide highly regular transport 
spectra, making the analysis of single\cite{nmat-cao:745} and multi quantum 
dot systems\cite{nnano-steele:363} possible. An unexpected feature in dc 
measurements was the observation of regions with suppressed or enhanced current 
level and abrupt, switching-like edges.\cite{strongcoupling, magdamping} These 
switching phenomena limit the ability to do finite bias spectroscopy at 
transparent tunnel barriers. They are consistent with feedback effects due to 
strong coupling between single electron tunneling and mechanical motion, as 
detailed in Ref. \onlinecite{Usmani2007}. The positive feedback between 
electronic tunneling and mechanical motion gives rise to self-oscillation of 
the vibration mode and in turn leads to an abrupt change in the current through 
and in the conductance of the quantum dot embedded on the mechanically active 
part of the carbon nanotube. 

In this manuscript we compare measurements on a highly stable carbon nanotube 
quantum dot device, which was subsequently cooled down first in the vacuum 
chamber of a conventional dilution refrigerator and afterwards within the 
$^{3}$He/$^{4}$He mixture (dilute phase) of a top-loading dilution refrigerator. 
A distinct difference between the measurement runs is that the abovementioned 
switching phenomena are absent when the sample is immersed into the viscous 
liquid $^{3}$He/$^{4}$He mixture. This provides additional support for their 
vibrational origin.

\begin{figure}[t]
\epsfig{file=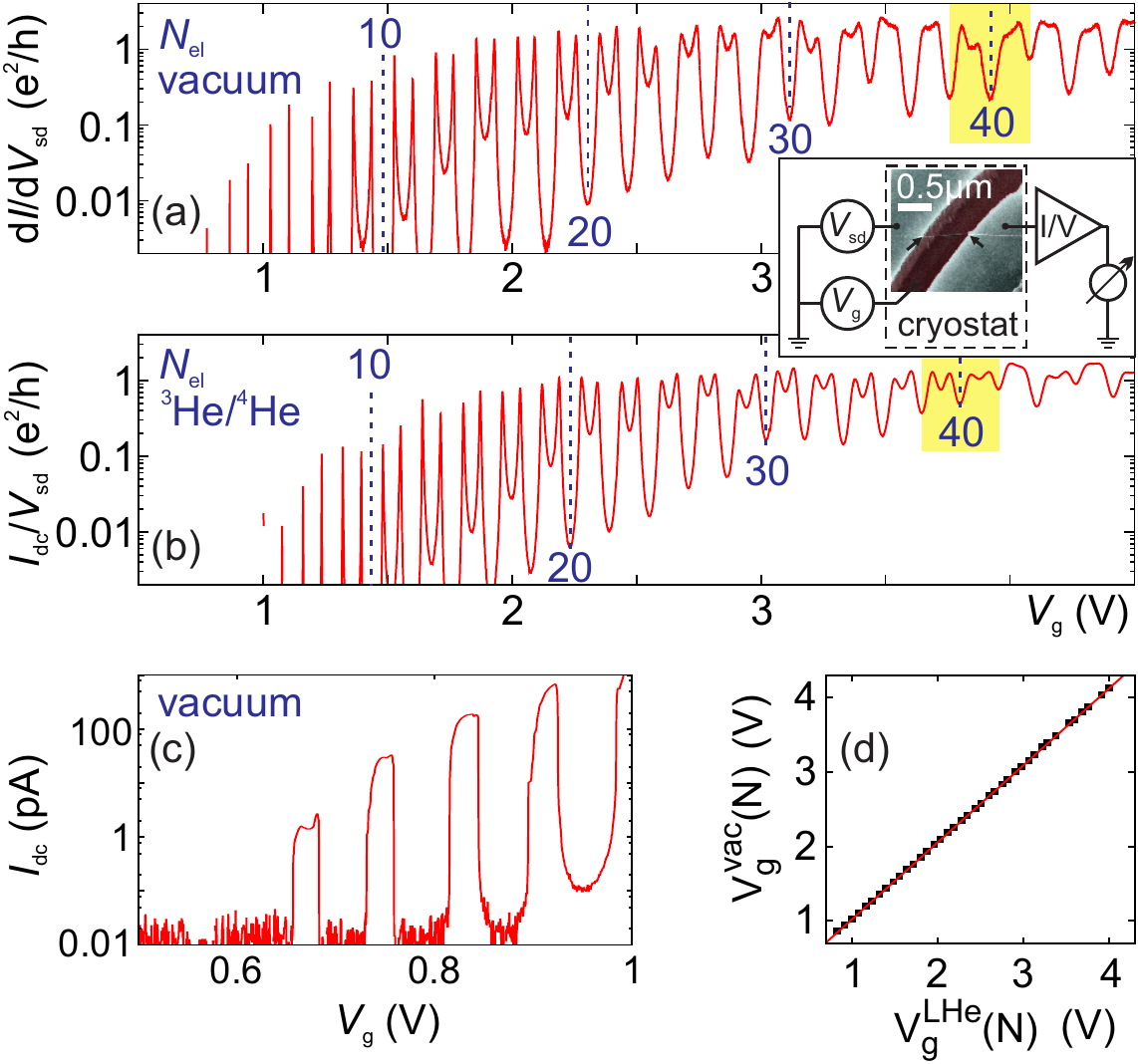,width=\columnwidth}
\caption{\label{fig-setup}
(Color online) (a) Low-bias differential conductance \didv\ of a 
carbon nanotube as a function of the gate voltage \vg, at dilution refrigerator 
base temperature in vacuum (lock-in measurement using $V_\text{sd,ac} = 
5\,\mu\text{V}$ at $f_\text{sd}=137.7\un{Hz}$). A Coulomb blockade dominated 
few electron regime is visible. For higher electron numbers \nel\ the tunnel 
rates gradually increase. Inset: Scanning electron micrograph of a similar 
carbon nanotube device with a simplified schematic of the measurement setup. 
(b) Conductance from a dc measurement 
$I_\text{dc}(\vg)$ at $V_\text{sd,dc} = 105\,\mu\text{V}$, subsequent cooldown 
in $^3$He/$^4$He mixture. In both cases, the gate voltage range shown in 
Fig.~\ref{fig-data} is shaded yellow. (c) Example measurement of the dc current 
at high bias $\vsd=12\un{mV}$, displaying the first Coulomb oscillations next 
to the nanotube band gap (vacuum case). (d) Single electron tunneling peak 
positions from (a) and (b). For each data point, the x-coordinate is given by 
the peak gate voltage in Helium (from (b)) and the y-coordinate by the 
corresponding peak gate voltage in vacuum (from (a)).}
\end{figure}
Our device consists of a clean carbon nanotube grown via chemical vapor 
deposition across a pre-fabricated trench and rhenium electrodes. Fabrication 
and structure details can be found in previous publications, as well as 
additional measurement data on the same device.\cite{magdamping, kondoso} 
Electronic transport measurements were performed subsequently in two dilution 
refrigerators at $25\un{mK} \lesssim T_\text{MC,base} \lesssim 30\un{mK}$. 
During the first of these cooldowns, the device was in vacuum environment, 
in the second cooldown using a top-loading dilution refrigerator the sample was 
mounted within the liquid $^3$He/$^4$He mixture (dilute phase) of the mixing 
chamber.

The inset of Fig.~\ref{fig-setup} shows a simplified sketch of the electronic 
measurement setup, as typically used in Coulomb blockade 
spectroscopy:\cite{kouwenhoven} a gate voltage \vg\ and a bias voltage \vsd\ 
are applied, and the resulting current is converted to a voltage at room 
temperature and recorded. As can be seen from conductance measurements as in 
Fig.~\ref{fig-setup}(a) and Fig.~\ref{fig-setup}(b), a small band gap separates 
the highly transparent hole conduction side (not shown here) from sharp Coulomb 
blockade oscillations in electron conduction. At higher electron number \nel\ 
the tunnel barriers to the contacts become increasingly transparent, and 
co-tunneling and regular Kondo enhancement of the conductance
emerge.\cite{nature-goldhaber:156, kondoso}

Fig.~\ref{fig-setup}(a) and Fig.~\ref{fig-setup}(b) compare the low-bias 
conductance of the device in the two subsequent cooldowns in vacuum 
(Fig.~\ref{fig-setup}(a)) and  $^{3}$He/$^{4}$He mixture 
(Fig.~\ref{fig-setup}(b)); a highly regular addition spectrum emerges in both 
cases. By recording current at increasingly large bias voltages and long 
amplifier integration times, as e.g. plotted in Fig.~\ref{fig-setup}(c), the 
first Coulomb oscillation next to the electronic band gap has been identified 
in both cases, providing the absolute number of trapped electrons. In liquid 
environment, the Coulomb blockade oscillations are shifted towards lower gate 
voltage; the gate voltage position of the Nth conductance maximum approximately 
follows a linear scaling $V_\text{g}^\text{vac}(N) \approx 1.037 \times 
V_\text{g}^\text{LHe}(N)-0.01\un{V}$, see also Fig.~\ref{fig-setup}(d). Such a 
scaling is already expected from the much-simplified model of a classical 
single electron transistor, see e.g. Ref.~\onlinecite{kouwenhoven}, where the 
distance between Coulomb oscillations $\Delta \vg$ is given by $\Delta 
\vg=e/C_g$. In our device, only part of the volume between nanotube and highly 
doped gate substrate is etched and thereby changes dielectric constant between 
cooldowns. Thus, the scaling of the gate capacitance $C_g$ is consistent with 
the upper bound given by the liquid helium dielectric constant\cite{Hunklinger} 
$\epsilon_{r,{}^4\text{He}} \approx 1.06$.

\begin{figure}[t]
\epsfig{file=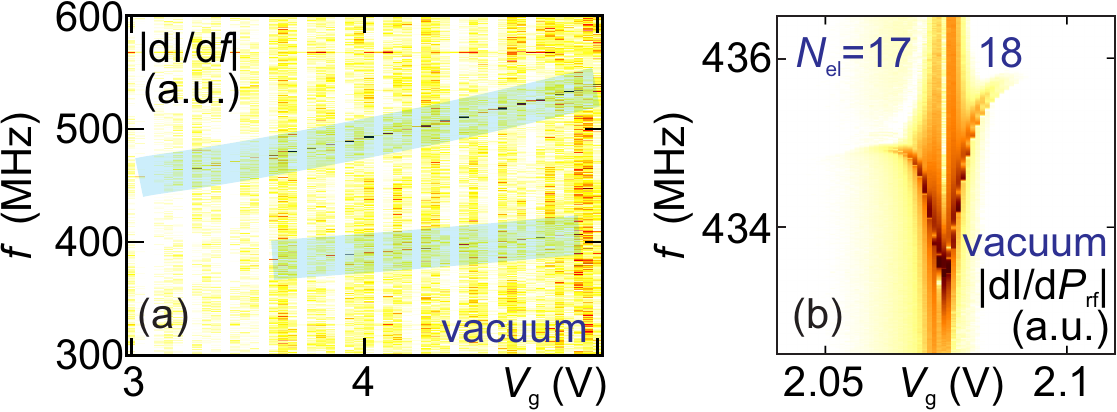,width=\columnwidth}
\caption{\label{fig-driven}
(Color online) (a) dc-detection of transversal mechanical resonances under 
contact-free rf irradiation, following Refs.~\onlinecite{highq, highqset}: 
$\left| \text{d}I_{\text{dc}} / \text{d}f\right|$ as function of gate voltage 
\vg\ and driving frequency $f$; two gate-dependent mechanical resonances are 
highlighted. (b) Exemplary detection of a mechanical resonance across one 
Coulomb oscillation, cf.\ Refs.~\onlinecite{strongcoupling, highqset, 
prb-meerwaldt:115454}. The rf source power is slowly modulated and the 
corresponding variation in current detected.
}
\end{figure}
In vacuum environment measurements, following Ref.~\onlinecite{highq}, the 
device displays under rf irradiation a clear resonant gate-dependent signal 
which can be attributed to the transversal vibration mode. This is illustrated 
in Fig.~\ref{fig-driven}(a) and Fig.~\ref{fig-driven}(b), displaying a large 
scale measurement\cite{highq} as well as an exemplary detail zoom into one 
Coulomb blockade oscillation.\cite{strongcoupling} Mechanical quality factors 
of $Q\simeq 25000$ have been observed; see Ref.~\onlinecite{magdamping} for a 
discussion of driven vibrations and the magnetic field dependence of mechanical 
effects.

In the following we focus on the intermediate coupling regime around $\nel=40$, 
shaded (yellow) in Fig.~\ref{fig-setup}(a,b), where the Kondo effect dominates 
the low-bias conductance in the Coulomb valleys with odd electron numbers. For a 
closer investigation the current is measured as function of gate voltage 
\vg\ and bias voltage \vsd, yielding the quantum dot stability diagram. The 
clean and essentially unchanged electronic structure allows us to compare 
measurements from two cool-downs but for the same quantum dot parameter regime. 

\begin{figure}[t]
\epsfig{file=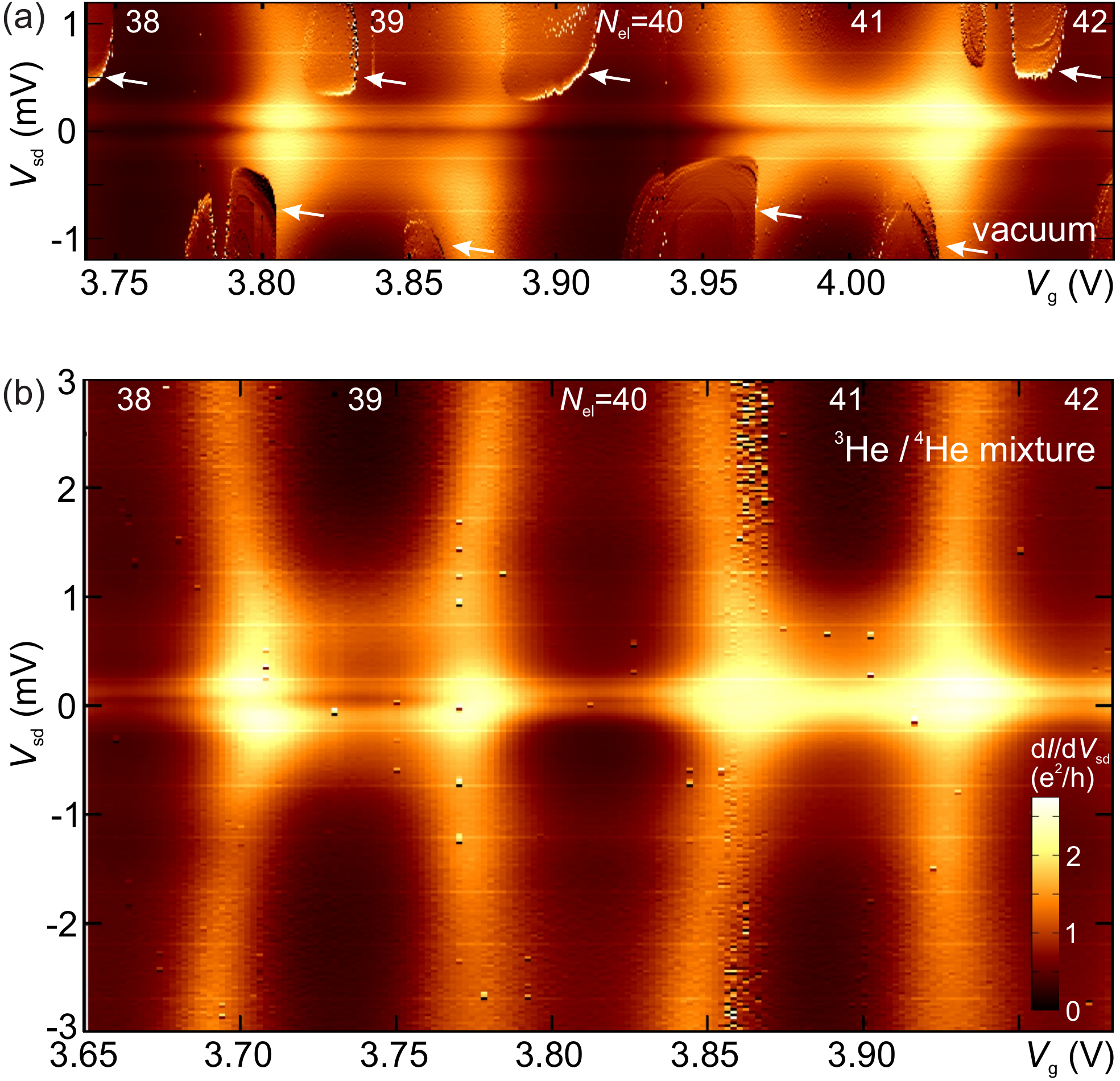,width=\columnwidth}
\caption{\label{fig-data} (Color online)
(a) Numerically derived differential conductance \didv\ as function of bias 
voltage \vsd\ and gate voltage \vg. The device is in vacuum. Kondo ridges of 
enhanced conductance\cite{nature-goldhaber:156, Nygaard2000, kondoso} are 
clearly visible; the dark line of lower conductance at $\vsd=0$ reflects the 
superconducting energy gap of the rhenium contacts. At finite bias, distinct 
switching phenomena due to electromechanical feedback\cite{strongcoupling, 
magdamping, Usmani2007} are clearly visible (white arrows). Related data has 
already been published in Ref.~\onlinecite{magdamping}, Figure 2. (b) 
Comparable measurement of the same sample, but now immersed into the liquid 
$^{3}$He/$^{4}$He mixture of a dilution refrigerator; same \vsd\ scaling. The 
switching phenomena are clearly absent even for significantly higher bias 
voltage \vsd.
}
\end{figure}
Figure~\ref{fig-data} contrasts those measurements and again demonstrates the 
consistency of the electronic features. The wide high conductance ridges at 
odd electon number \nel\ are a manifestation of the Kondo 
effect;\cite{nature-goldhaber:156, Nygaard2000, kondoso} its strong emergence 
at 
odd electron numbers is already clearly visible in Fig.~\ref{fig-setup}(a,b) 
for $N>10$. The overall reduced conductance closely around $\vsd=0$ accompanied 
by two lines of enhanced conductance at low finite \vsd\ can be tentatively 
attributed to the presence of an energy gap $\Delta$ in the superconducting 
rhenium contacts\cite{inelasticcopenhagen, nbcot} and is also similar in both 
measurements. At a finite voltage of $e\vsd=\pm 2\Delta$ the onset of 
quasiparticle cotunneling leads to additional conductance.

While in Fig.~\ref{fig-data}(a) the sample is in vacuum, in (b) the sample is 
immersed into liquid $^3$He/$^4$He mixture (diluted phase). A significant 
difference between the two data sets is given by the sharp spikes in the 
conductance in vacuum, Fig.~\ref{fig-data}(a), highlighted with white arrows, 
framing regions of enhanced or suppressed current level. This phenomenon is well 
understood by a positive feedback between single electron tunneling and the 
mechanical motion.\cite{Usmani2007} It requires large tunnel rates between 
quantum dot and leads; first traces of the effect are already visible at much 
lower electron number (see also Ref.~\onlinecite{kondoso} Fig.~11). In the 
measurement of Fig.~\ref{fig-data}(b), plotted at identical y-axis scaling, 
these feedback-induced phenomena are clearly absent, even up to a much higher 
bias voltage.\cite{fluctuation}

While electronic damping mechanisms have been demonstrated 
recently,\cite{magdamping, prb-meerwaldt:115454} here the viscous 
$^3$He/$^4$He fluid surrounding the carbon nanotube adds purely mechanical 
damping, reduces the high quality factor of the mechanical system, and 
suppresses  self-oscillation in the system. This provides a much more direct 
confirmation that mechanical motion is the underlying mechanism for the 
switching phenomena.

Mechanical resonators have long been used for viscosity measurements in 
low-temperature physics, and recently also nanomechanical systems have entered 
this field.\cite{nanotech-kraus:165, jltp-collin:653, rsi-gonzalez:025003} The 
unusual parameter combination present in our setup, combining a $^3$He/$^4$He 
fluid at millikelvin temperatures with a beam diameter on the order $\sim 
1\un{nm}$ and resonance frequencies $\sim 250\un{MHz}$,\cite{magdamping} poses 
however challenges for a detailed estimate on the damping mechanisms. The 
viscosity of the diluted phase of a $^3$He/$^4$He mixture at $T \simeq 
25\un{mK}$ is $\eta\sim 10^{-4}\un{N\,s/m}^2$.\cite{Hunklinger} This is 
roughly an order of magnitude larger than the viscosity of air at room 
temperature and standard condition in pressure, $\eta\sim 
10^{-5}\un{N\,s/m}^2$. Since at room temperature already a much smaller gas 
pressure of $p\sim 1000\un{Pa}$ causes a reduction of the effective mechanical 
quality factor of carbon nanotube resonators to $Q \ll 
100$,\cite{nature-sazonova:284, jjap-fukami:06fg04} a strong mechanical damping 
by the viscous medium is expected. 

Apart from damping of the vibrational mode, in addition, the $^3$He/$^4$He 
liquid naturally leads to improved thermalization of the entire chip structure 
and the cables connecting to it. However, the transversal vibration quality 
factor of a carbon nanotube device increases at lower temperature\cite{highq} 
-- which, as dominant effect, would enhance switching effects rather than 
suppressing them.\cite{strongcoupling, magdamping, Usmani2007}

In conclusion, we give an intuitive demonstration of the mechanical origin of 
switching phenomena\cite{strongcoupling, magdamping, Usmani2007} observed in 
freely suspended carbon nanotubes by comparison of measurements in different 
sample environments. Keeping the electronic properties almost unaffected, a 
viscous $^3$He/$^4$He medium supplies an additional damping mechanism and 
prevents feedback-induced oscillation. Hence, this damping mechanism 
provides a way to circumvent instabilities in finite bias spectroscopy on 
overgrown freely suspended carbon nanotubes without restriction of the 
accessible temperature, magnetic field, and tunnel coupling range.

\begin{acknowledgments}
The authors acknowledge financial support by the Deutsche Forschungsgemeinschaft
(Emmy Noether grant Hu 1808/1, GRK 1570, SFB 631 TP A11) and by the 
Studienstiftung des deutschen Volkes. We thank A.\ Dirnaichner for experimental 
support.
\end{acknowledgments}

\bibliography{paper}

\begin{thebibliography}{29}
\bibitem [{flu()}]{fluctuation}%
  \BibitemOpen
  \href@noop {} {\ }\bibinfo {note} {A distinct fluctuation is visible in
  Fig.~\ref{fig-data}(b) at $\vg=3.86\un{V}$. Similar phenomena occur at
  $\vg=3.44\un{V}$ and $\vg=4.25\un{V}$. They display only very weak dependence
  on bias voltage over the measured range $-8\un{mV}\le\vsd\le 8\un{mV}$, and
  their gate voltage position does not stand in any clear relation to the
  position of the Coluomb blockade oscillations. Both these observations
  indicate that a charge trap unrelated but close to the quantum dot at hand is
  triggered.}\BibitemShut {Stop}%
\end{thebibliography}

\end{document}